\documentclass[sigconf,natbib=true]{acmart}

\AtBeginDocument{%
  \providecommand\BibTeX{{%
    \normalfont B\kern-0.5em{\scshape i\kern-0.25em b}\kern-0.8em\TeX}}}

\setcopyright{acmlicensed}
\copyrightyear{2024}
\acmYear{2024}
\acmDOI{10.1145/3626772.3657838}

\acmConference[SIGIR '24] {Proceedings of the 47th International ACM SIGIR Conference on Research and Development in Information Retrieval}{July 14--18, 2024}{Washington, DC, USA.}
\acmBooktitle{Proceedings of the 47th International ACM SIGIR Conference on Research and Development in Information Retrieval (SIGIR '24), July 14--18, 2024, Washington, DC, USA}
\acmISBN{979-8-4007-0431-4/24/07}

\acmSubmissionID{9954}

\usepackage[ruled,linesnumbered]{algorithm2e}

\usepackage{dsfont}
\usepackage{multirow}
\usepackage{amsmath}
\usepackage{enumitem}
\usepackage{eucal}
\usepackage{bm}
\usepackage{amsfonts}
\usepackage{subfigure}
\usepackage[switch]{lineno}
\usepackage{bbding}
\usepackage{listings}
\usepackage{balance}

\begin{document}

\title{Contrast then Memorize: Semantic Neighbor Retrieval-Enhanced \\ Inductive Multimodal Knowledge Graph Completion}


\author{Yu Zhao}
\email{zhaoyu@dbis.nankai.edu.cn}
\affiliation{%
  \institution{College of Computer Science\\VCIP, TMCC, TBI Center\\Nankai University}
  \city{Tianjin}
  \country{China}
}

\author{Ying Zhang}
\authornote{Corresponding author.}
\email{yingzhang@nankai.edu.cn}
\affiliation{%
  \institution{College of Computer Science\\VCIP, TMCC, TBI Center\\Nankai University}
  \city{Tianjin}
  \country{China}
}

\author{Baohang Zhou}
\email{zhoubaohang@dbis.nankai.edu.cn}
\affiliation{%
  \institution{College of Computer Science\\VCIP, TMCC, TBI Center\\Nankai University}
  \city{Tianjin}
  \country{China}
}

\author{Xinying Qian}
\email{qianxinying@dbis.nankai.edu.cn}
\affiliation{%
  \institution{College of Computer Science\\VCIP, TMCC, TBI Center\\Nankai University}
  \city{Tianjin}
  \country{China}
}

\author{Kehui Song}
\email{kehui.song10@gmail.com}
\affiliation{%
  \institution{School of Software\\Tiangong University}
  \city{Tianjin}
  \country{China}
}

\author{Xiangrui Cai}
\email{caixr@nankai.edu.cn}
\affiliation{%
  \institution{College of Computer Science\\VCIP, TMCC, TBI Center\\Nankai University}
  \city{Tianjin}
  \country{China}
}

\renewcommand{\shortauthors}{Yu Zhao, et al.}

\begin{abstract}
A large number of studies have emerged for Multimodal Knowledge Graph Completion (MKGC) to predict the missing links in MKGs.
However, fewer studies have been proposed to study the inductive MKGC (IMKGC) involving emerging entities \textit{unseen during training}. 
Existing inductive approaches focus on learning textual entity representations, which neglect rich semantic information in visual modality.
Moreover, they focus on aggregating structural neighbors from existing KGs, which of emerging entities are usually limited.
However, the semantic neighbors are decoupled from the topology linkage and usually imply the true target entity.
In this paper, we propose the IMKGC task and a semantic neighbor retrieval-enhanced IMKGC framework CMR, where the \textit{contrast} brings the helpful semantic neighbors close, and then the \textit{memorize} supports semantic neighbor retrieval to enhance inference.
Specifically, we first propose a unified cross-modal contrastive learning to simultaneously capture the textual-visual and textual-textual correlations of query-entity pairs in a unified representation space. 
The contrastive learning increases the similarity of positive query-entity pairs, therefore making the representations of helpful semantic neighbors close.
Then, we explicitly memorize the knowledge representations to support the semantic neighbor retrieval. 
At test time, we retrieve the nearest semantic neighbors and interpolate them to the query-entity similarity distribution to augment the final prediction.
Extensive experiments validate the effectiveness of CMR on three inductive MKGC datasets.
Codes are available at https://github.com/OreOZhao/CMR.

\end{abstract}






\maketitle

\section{Introduction}

\begin{figure}[t]
\centering
\subfigure[IMKGC aims to predict missing entity of query (\textit{WALL-E}, \textit{production companies}) where \textit{WALL-E} is unseen during training. The semantic neighbors of the query could help with predicting \textit{Pixar}.]
{
\label{fig:intro_knn}
\includegraphics[width=0.48\textwidth]{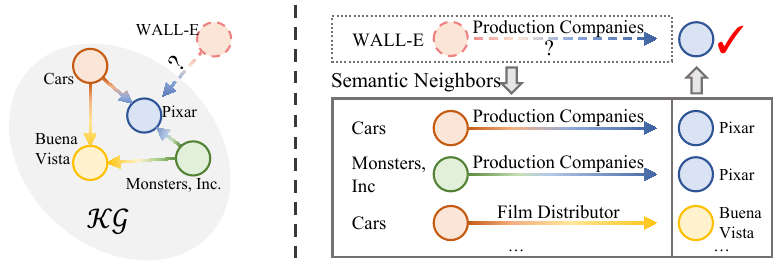}}
\subfigure[Multimodal semantic correlation (shading) between query and its target entity.]{
\label{fig:intro_clip}
\vspace{-25pt}
\includegraphics[width=0.48\textwidth]{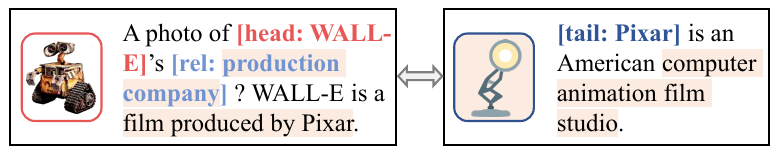}}
\vspace{-15pt}
\caption{We propose to aggregate semantic neighbors to augment inductive MKGC (Figure \ref{fig:intro_knn}). 
Moreover, we propose to capture both textual-textual and textual-visual correlation in the query-entity pairs (Figure \ref{fig:intro_clip}).}
\vspace{-15pt}
\end{figure}

Multimodal Knowledge Graphs (MKGs) \cite{Liangke2022mmkgSurvey,zhu2022mmkgsurvey} organize multimodal factual knowledge composed of entities and relations and have been widely applied to many knowledge-driven tasks \cite{sun2020mmkgrs,marino2019okvqa}.
Due to the incompleteness of KGs, the Multimodal Knowledge Graph Completion (MKGC) task \cite{xie2017IKRL,xie2016DKRL} has been proposed to predict the missing entities in KGs. For a query $q=(h,r,?)$ involving the head entity and relation, MKGC aims to predict the tail entity $t$ with multimodal information, i.e. structure, text, and visual modalities.
However, in the real world, new entities emerge through time and propose challenges for the MKGC model to be inductive and generalize to unseen entities without re-training.
Therefore, we propose the Inductive Multimodal Knowledge Graph Completion (IMKGC) task, aiming to complete missing links involving unseen entities with multimodal information.
As shown in Figure \ref{fig:intro_knn}, \textit{WALL-E} is unseen during training and IMKGC aims to predict its \textit{production companies}.

To incorporate multimodal information, many MKGC methods have been proposed based on 
knowledge graph embedding (KGE) methods \cite{li2023imf,cao2022otkge,zhao2022mose}, or pre-trained vision-language models (VLMs) \cite{chen2022mkgformer,liang2023SGMPT}. 
However, they are not compatible with the inductive setting: the entity representation table of KGE-based methods and entity tokens of VLM-based methods all require re-training to generalize to emerging entities. 
The way to generalize the MKGC model to unseen entities without re-training is less explored.

Although there have been inductive KGC studies \cite{markowitz2022statik,daza2021blp,wang2021StAR} to predict the missing links involving unseen entities, they still have two main issues: 
\textbf{(1) They neglected that visual modality could help with inductive inference.} 
Existing inductive approaches learn the textual entity embedding based on pre-trained language models (PLMs) \cite{devlin2019bert}.
However, with the help of pre-trained visual models \cite{radford2021CLIP,dosovitskiy2020ViT},
the visual representations could also transfer to emerging entities that are not seen during training.
Moreover, the semantic correlation between entity images and queries could indicate the true entity.
As shown in Figure \ref{fig:intro_clip}, the texts ``\textit{production company}'', ``\textit{produced by Pixar}'' from the query are not only related to the descriptions of \textit{Pixar} but also corresponded to the image of \textit{Pixar} logo. 
\textbf{(2)
The helpful semantic neighbors are not explicitly aggregated.} 
Existing inductive approaches only aggregate structural information, which of the emerging entity is usually limited.
The semantic neighbors, which \textit{share similar semantics with the query and are not necessarily structurally linked}, usually imply the true target entity. 
As shown in Figure \ref{fig:intro_knn}, the semantic neighbors of the query (\textit{WALL-E}, \textit{production companies}), 
such as \textit{(Cars, production companies)} with similar semantics, directly indicate the target entity \textit{Pixar}.
The aggregation from such helpful semantic neighbors could help with inductive inference.

Though incorporating visual modality and semantic neighbors is intuitive for IMKGC, there still are several challenges:
\textbf{(1) The multimodal contradiction.} As shown in Figure \ref{fig:intro_clip}, the images of the head entity and tail entity in a triple are usually dissimilar, which contradicts the correlation in text modality.
\textbf{(2) Semantic neighbor representation learning}. 
Unlike the structural neighbor sampling by topology linkage, the semantic neighbors are sampled by representation similarity. 
The helpful semantic neighbors with the same target could only be sampled to help the prediction if their representations are similar to the query. 
However, it is computationally expensive to train the representations of helpful semantic neighbors closer to the query than all other queries.


In this paper, we propose to address the challenges with \textit{contrast then memorize} strategy, in which the \textit{contrast} not only avoids the modality contradiction but also brings the helpful semantic neighbors together, and the \textit{memorize} supports the semantic neighbor sampling and aggregation at test time.
We propose a novel IMKGC framework \textbf{CMR} including three steps of \textbf{C}ontrastive learning, \textbf{M}emorization, and \textbf{R}etrieval. 
Specifically, (1) \textbf{unified cross-modal contrastive learning} captures the textual-textual and textual-visual correlations of the positive query-entity pairs in a unified representation space, which prevents contradiction from the visual-visual dissimilarity. 
We employ a query-entity bi-encoder architecture, where the entity encoder is equipped with a visual mapping network to project the images into trainable prefixes in textual space.
The contrastive learning optimizes the encoders to increase the similarity between positive query-entity pairs, which thereby makes the representations of helpful semantic neighbors with the same target entity close.
\textbf{(2) Knowledge representation memorization} explicitly memorizes the query embeddings in a knowledge store after training. 
Like the KG supports structural neighbor sampling, the knowledge store supports semantic neighbor sampling with the help of search tools \cite{johnson2019faiss}.
\textbf{(3) Retrieval-augmented inference} linearly interpolates the query-entity similarity distribution with the retrieved semantic neighbors from the knowledge store at test time, which benefits from our proposed contrastive learning and avoids expensive semantic neighbor training.
Experiments on three IMKGC datasets and three transductive MKGC datasets validate the effectiveness and generalizability of CMR.
The contributions of the paper can be summarized as follows:
\begin{itemize}
    \item To the best of our knowledge, we are the first to study the IMKGC to address the emerging entity generalization problem of MKGs. 
    We propose to enhance IMKGC with textual-visual correlations and semantic neighbors.
    \item We propose a framework CMR: (1) \textbf{contrastive learning} avoids modality contradiction and brings helpful semantic neighbors close, (2) \textbf{memorization} explicitly stores semantic neighbors, and (3) \textbf{retrieval} aggregates the semantic neighbors at test time to augment inference.
    \item Extensive experiments demonstrate the effectiveness of CMR on both inductive and transductive datasets. Experiments also show that the proposed unified cross-modal contrastive learning and semantic neighbors retrieval could help with generalization to emerging entities.
\end{itemize}

\section{Related Work}

\subsection{Multimodal Knowledge Graph Completion}
MKGC \cite{xie2017IKRL,xie2016DKRL} aims to utilize auxiliary multimodal information such as texts and images to complement the structure information for missing entity prediction. 
In our paper, we focus on the inductive MKGC involving emerging entities unseen during training.

Some MKGC methods augment the single-modal KGE methods \cite{bordes2013TransE, lacroix2018complexn3} to multimodal information \cite{pezeshkpour2018mkbe,wang2019transAE,wang2021rsme,zhao2022mose,cao2022otkge,li2023imf}, whose representation table could not generalize to unseen entities.
Recently, some methods \cite{chen2022mkgformer,liang2023SGMPT} based on pre-trained VLMs transform the entity as additional tokens and perform MKGC as Masked Language Modeling (MLM) problem. 
However, the token embeddings of unseen entities are randomly initialized without training, thus it is hard for MLM to predict unseen entities.
In our paper, we focus on optimizing encoders with contrastive learning and augmenting query-entity similarity inference with semantic neighbor retrieval instead of with MLM prediction to address the unseen entity generalization problem.

Moreover, some methods point out the multimodal contradiction problem \cite{wang2021rsme,zhao2022mose,chen2022mkgformer} that the images of head and tail entities show low relevance.
RSME \cite{wang2021rsme} proposes to filter out noisy visual information with a filter gate.
MoSE \cite{zhao2022mose} proposes to split the multimodal relation representations while ensemble multimodal prediction to alleviate inter-modal interference.
MKGformer \cite{chen2022mkgformer} proposes a multi-level multimodal fusion mechanism to alleviate noisy information.
In our paper, we propose to capture the textual-textual and textual-visual correlation between queries and entities with unified cross-modal contrastive learning to prevent interference from visual-visual dissimilarity.

\subsection{Inductive Knowledge Graph Completion}
There have been developing focuses on the IKGC problem to address the unseen entity problem.
Some works focus on learning unseen entity representation by aggregating the structural neighbors to existing KG \cite{wang2019LAN,dai2020InvTransE}.
Recently, with the development of PLMs such as BERT \cite{devlin2019bert}, 
some works propose to utilize transferable text information.
They encode the query \cite{wang2021StAR} or entity \cite{daza2021blp,markowitz2022statik} to text embedding with BERT. 
Then they learn the structural information by fine-tuning BERT after structural objectives \cite{daza2021blp,wang2021StAR} such as TransE \cite{bordes2013TransE} or aggregating subgraph neighbors \cite{markowitz2022statik} through GNNs \cite{gilmer2017MPNN}.
However, none of these methods explore entity images in learning generalizable entity representations. Moreover, emerging entities have fewer related triples thus the structural information is limited for inference. 
In our paper, we propose to utilize both visual and textual information for inductive knowledge representation learning.
We also propose semantic neighbor retrieval-enhanced inference, which frees the model from structural linkage and involves semantic-similar but not necessarily structural-linked entities.

\subsection{Retrieval-augmented Model for KG }
Retrieval-augmented language models \cite{khandelwal2019KNNLM} aims to improve the model performance by explicitly memorizing the patterns in a datastore rather than implicitly memorizing them in model parameters for better generalization. 
Inspired by them, many studies utilize retrieval-augmented methods for knowledge graph construction tasks, such as relation extraction \cite{chen2022REopenbook}, named entity recognition \cite{wang2022knnner}, knowledge extraction \cite{yao2023schema}.

Recently, $k$NN-KGE \cite{zhang2022kNNKGE} proposes retrieval-augmented MLM entity prediction for KGC.
However, the generalization of MLM entity prediction is limited and the knowledge representations are not explicitly trained to enhance the retrieval performance.
In our paper, unlike only improving generalization through memorization \cite{khandelwal2019KNNLM,zhang2022kNNKGE}, we propose contrastive learning before memorization to increase the representation similarity of semantic neighbors with the same target to improve the retrieval-augmented inference.

\section{Preliminaries}
\subsection{Multimodal Knowledge Graphs} 
Multimodal knowledge graphs (MKGs) are knowledge graphs with multimodal information, such as images and texts.
A KG could be denoted as $\mathcal{G} = \{\mathcal{E}, \mathcal{R}, \mathcal{T} \}$, where $\mathcal{E}$ is the set of entities, $\mathcal{R}$ is the set of relations, $\mathcal{T}$ is the set of triples $\{(h,r,t)\}\subseteq \mathcal{E}\times\mathcal{R}\times\mathcal{E}$. 
For MKG, each entity $e \in \mathcal{E}$ has multimodal semantics from descriptions $e_d$ and images $e_v$, and each relation $r \in \mathcal{R}$ has descriptions $r_d$.

\subsection{Multimodal Knowledge Graph Completion}
For a particular query $q = (h,r,?)$, Multimodal Knowledge Graph Completion (MKGC) aims to predict the missing entity with multimodal information, which is ranking all the candidate entities $e\in\mathcal{E}$ and making the target entity $t$ the highest. The reversed triple $(t, r^{-1}, h)$ are also added for unifying the head prediction and tail prediction into tail prediction \cite{bordes2013TransE}.

\subsubsection{Transductive MKGC} 
Under the transductive setting, the entity set in the train set and the test set are both $\mathcal{E}$, meaning all the entities in the test set were seen in the training phase.

\subsubsection{Inductive MKGC}
Under the inductive setting, emerging entities appear in the test time, which are unseen in the training phase.
We define the set of seen entities as $\hat{\mathcal{E}}$, the set of unseen entities as $\tilde{\mathcal{E}}$, and $\hat{\mathcal{E}} \cup \tilde{\mathcal{E}} = \mathcal{E},\hat{\mathcal{E}}\cap \tilde{\mathcal{E}}=\emptyset $.
The triples in the test set all involve unseen entities as 
$\tilde{\mathcal{T}}=\{(\tilde{e},r,\hat{e})~or~ (\hat{e},r,\tilde{e})| \tilde{e}\in\tilde{\mathcal{E}}, \hat{e} \in \hat{\mathcal{E}}, r\in\mathcal{R} \}$.
The IMKGC model ranks both seen and unseen entities $e\in\mathcal{E}$ for predicting the target entity.
We do not assume unseen relation \cite{daza2021blp}.

\section{Methodology}

\begin{figure*}[!t]
    \centering
    \includegraphics[width=1\textwidth]{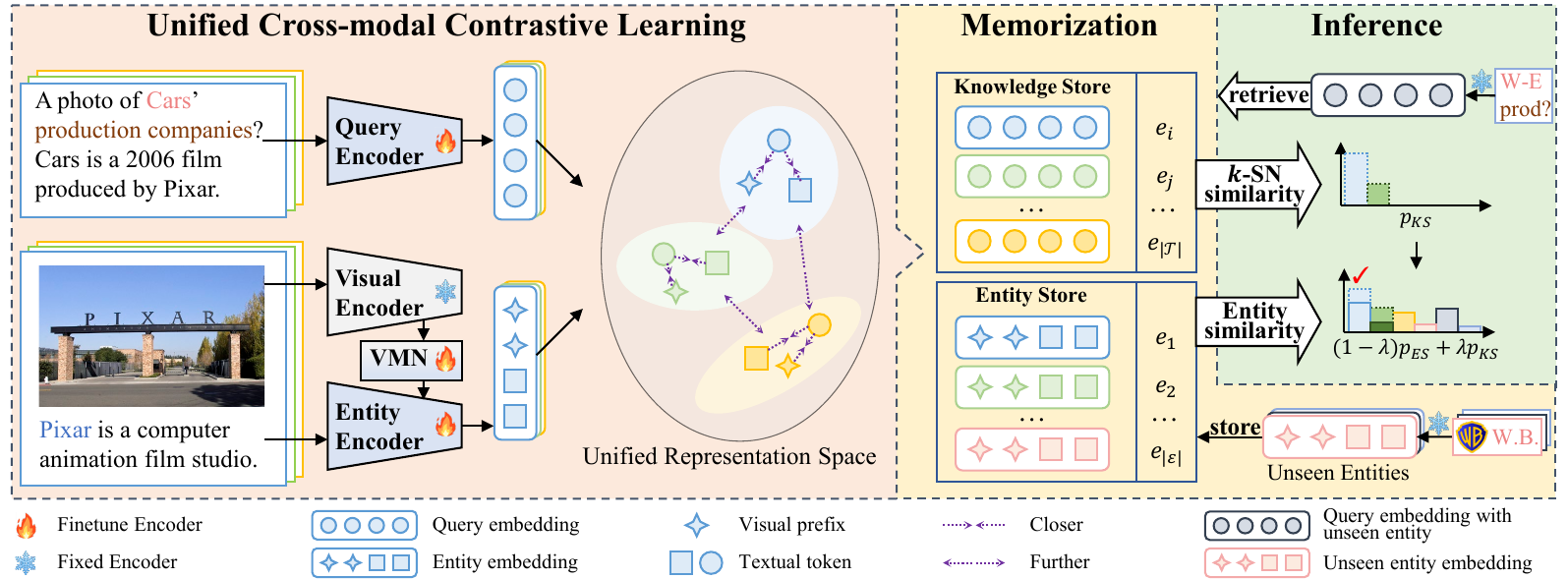}
    \caption{Contrastive learning, Memorization, and Retrieval (CMR) framework for Inductive Multimodal Knowledge Graph Completion. Contrastive learning optimizes the bi-encoders to shorten the distance between query-entity pairs and to capture multimodal semantic correlation in a unified representation space. Memorization explicitly memorizes the knowledge representations after training. The retrieval from the knowledge store aggregates the semantic neighbors and interpolates them to the query-entity similarity distribution for the final prediction at test time.}
    \label{fig: model_overview}
\end{figure*}

\subsection{Overview}

The CMR framework is shown in Figure \ref{fig: model_overview}. 
First, we reformulate the triples $(h,r,t)$ in the train graph as prompts from queries $q=(h,r,?)$ and entities $t$.
We utilize a bi-encoder architecture equipped with a visual mapping network (VMN) to encode the queries and entities. 
We propose a unified cross-modal contrastive knowledge representation learning to shorten the distance between textual queries and multimodal entities to capture the textual-textual and textual-visual correlations of query-entity pairs in a unified representation space.
After the contrastive learning, the semantic neighbors with the same target entity are also closer.

Then we construct a knowledge store (KS) that memorizes the query embeddings and their target entity labels. 
We also construct an entity store (ES) with entity embeddings with all the entities.
For inductive inference, with a query $q=(h,r,?)$, we first retrieve its $k$-nearest semantic neighbor queries in the KS according to representation distance.
The distribution based on negative distance forms the probability of KS $p_{KS}$.
Moreover, we compute the distance between the query and all the entities in the entity store to form the probability of ES $p_{ES}$.
We linearly interpolate the $p_{KS}$ with $p_{ES}$ to produce the final prediction.

\subsection{Unified Cross-modal Contrastive Learning}
As stated by the previous studies in MKGs \cite{zhao2022mose,chen2022mkgformer,wang2021rsme}, the multimodal contradictions hinder the MKG representation learning, especially with the images between the head entity and tail entity not semantically similar. 
To effectively incorporate the visual modality, 
we propose to incorporate the visual modality by learning the cross-modal interactions between the textual queries and visual-textual target entities.

\subsubsection{Query Encoder} \label{sec: query encoder}
The query encoder is a PLM used for encoding the texts of query $q=(h,r,?)$.
The input of the query encoder is the prompt that combines the descriptions of head entity $h_d$ and relation $r_d$. 
The template of query is denoted as Equation \eqref{eq: template}, where $h$ is the name of head entity.
\begin{equation} \label{eq: template}
    T_q(h_d, r_d)=\text{``}\texttt{[CLS] }\textit{A photo of }h\textit{'s } r_d\textit{? } \texttt{[SEP]} h_d\text{''}
\end{equation}
We define the query encoder as $f_q(\cdot, \cdot)$ and the output is the L2-normalized mean aggregation over all the last-layer hidden states of the query encoder. 
The output of the query encoder is the query embedding, denoted as Equation \eqref{eq: query emb}.
\begin{equation} \label{eq: query emb}
    \mathbf{q} = f_q(h, r) = f_q(T_q(h_d,r_d))
\end{equation}

\subsubsection{Visual Prefix-based Entity Encoder}
The entity encoder is a PLM used for encoding the target entity of a triple. 
In order to simultaneously capture the textual-textual and textual-visual correlations, 
inspired by the recent progress of visual prompt \cite{li2023acl-ModCR,liang2022promptfusion} and prefix-tuning \cite{li-liang-2021-prefixtuning}, we propose to convert the entity image representation from visual encoder into learnable visual prefixes in textual space.
In this way, the textual tokens and their visual prefixes could interact inside the PLMs.
To alleviate the modality gap from visual to text, we employ a visual mapping network (VMN) \cite{mokady2021clipcap} between the visual encoder and the entity encoder. 
Then the visual prefixes and the entity description are fed into the PLM entity encoder to obtain the entity representations.

The image feature from visual encoder is $\mathbf{v}_e = f_v(e_v)$. 
The VMN projects the image feature into a $l$-length sequence of visual prefixes, denoted as Equation \eqref{eq: vmn}. 
We adopt a two-layer MLP activated with ReLU as VMN.
Then, we concatenate the obtained visual prefixes embeddings to the token embeddings of entity descriptions as $[\mathbf{V}, \mathbf{D}]=(p_1, ..., p_l, d_{l+1}, ..., d_L)$, where $L$ is the length of the sequence.
The concatenation is fed into the PLM of the entity encoder. 
\begin{equation}\label{eq: vmn}
    \mathbf{V}=(p_1, ..., p_l)=\text{VMN}(\mathbf{v}_e)=\text{VMN}(f_v(e_v))
\end{equation}

We also use the L2-normalized mean aggregation over all the last-layer hidden states of the entity encoder as entity embedding. 
We denote the multimodal fusion entity representation from the entity encoder as Equation \eqref{eq: ent encoder}. Since we utilize mean aggregation, the fusion entity representation could be denoted as the sum of the visual component and the textual component.
\begin{equation} \label{eq: ent encoder}
\begin{aligned}
        \mathbf{e}_f & = f_e(e) = f_e([\mathbf{V},\mathbf{D}]) =\mathbf{e}_v + \mathbf{e}_d \\
\end{aligned}
\end{equation}

\subsubsection{Training Objectives}
The training objective of CMR is to optimize the encoders to increase the representation similarity between the query and its target entity in the positive triple $(h,r,t)$ in the train graph 
so that the encoders could generalize to the emerging entity in the test phase.
Moreover, the representations of the semantic neighbors with the same target entity would be close.

\textbf{Fusion contrastive loss.}
Since the entity set $\mathcal{E}$ is usually too large to put in a batch, we utilize the contrastive loss InfoNCE \cite{oord2018infonce} to increase the similarity between positive query-entity pairs and decrease that between the negative pairs.
We first propose a fusion contrastive loss $\mathcal{L}_{FC}$.
We denote the similarity function between query embedding and entity embedding as $s=\exp(\mathbf{q} \cdot \mathbf{e} /\tau)$, where $\tau$ is a temperature hyper-parameter that adjusts the importance of negative pairs. Since the embeddings are L2-normalized, the dot product equals the cosine similarity metric. 
The fusion contrastive loss for the $i$-th triple is calculated as Equation \eqref{eq: fc loss}, where $\mathcal{N}^i$ is the set of negative samples for the $i$-th triple.
\begin{equation} \label{eq: fc loss}
\begin{aligned}
    \mathcal{L}^i_{FC} &= - log \frac{s_f^i}{s_f^i + \sum_{n\in\mathcal{N}_i}s_f^n}  \\
    & = -log \frac{\prod_{m\in\mathcal{M}}s_m^i}{\prod_{m\in\mathcal{M}}s_m^i + \sum_{n\in\mathcal{N}^i}s_f^n} \\
\end{aligned}
\end{equation}

\textbf{\textit{Remark 1}}: 
\textit{Here we analyze how the $\mathcal{L}_{FC}$ handles the hard positive samples that have contradictory similarity values between modalities.
The similarity between query and fusion embedding could be seen as the product of all the similarities of multiple modalities as Equation \eqref{eq: sim_fusion}, where we denote $\mathcal{M}$ as a modality set without loss of generality for more modalities.}
\begin{equation} \label{eq: sim_fusion}
s_f=\exp(\mathbf{q}\cdot(\mathbf{e}_v+\mathbf{e}_d)/\tau)= s_v\cdot s_d =\prod_{m\in\mathcal{M}}{s_m}, \mathcal{M} = \{\mathcal{V}, \mathcal{D}\}
\end{equation}

\textit{The hard positive samples are the ones that have at least one hard modality $h\in \mathcal{M}$ that $s_h$ is small. 
For the modality $h$, the gradient from $\mathcal{L}_{FC}^i$ to $s_h^i$ is calculated as Equation \eqref{eq: fc loss partial}. 
Therefore, even if only one of the modalities has a small similarity, the hard positive samples can be trained with large updates from small $s_h^i$ and $s_f^i$ in the denominator.}
\begin{equation} \label{eq: fc loss partial}
\begin{aligned}
    \frac{\partial{\mathcal{L}^i_{FC}}}{\partial{s_h^i}} &= -\frac{\sum_{n\in\mathcal{N}_i}s_f^n}{s_h^i(\prod_{m\in\mathcal{M}}s_m^i+\sum_{n\in\mathcal{N}_i}s_f^n)} \\
    & = - \frac{\sum_{n\in\mathcal{N}_i}s_f^n}{s_h^i(s_f^i + \sum_{n\in\mathcal{N}^i}s_f^n)}
\end{aligned}
\end{equation}

\textbf{Pre-align contrastive loss. }
We propose a pre-align contrastive loss $\mathcal{L}_{AC}$ as Equation \eqref{eq: ac loss} for guiding the VMN module to project the image features into more accurate visual prefixes. 
The similarity between query embedding and the projected prefixes is denoted as $s_p=\exp(\mathbf{q}\cdot \bar{\mathbf{V}}/\tau)$ where $\bar{\mathbf{V}}$ is the mean pooling of visual prefix embeddings. 
The pre-align contrastive loss $\mathcal{L}_{AC}$ captures the textual-visual correlation between queries and entities. 
It further enhances the fusion contrastive loss $\mathcal{L}_{FC}$ as a middle-step constraint and improves the final prediction.
\begin{equation} \label{eq: ac loss}
    \mathcal{L}^i_{AC} = - log \frac{s_p^i}{s_p^i+ \sum_{n\in\mathcal{N}^i}s_p^n}
\end{equation}

\textbf{Overall objectives.} In addition to the query-to-entity losses, we consider bidirectional learning objectives and add the reversed entity-to-query losses $\mathcal{L}_{FC'}$ and $\mathcal{L}_{AC'}$.  
For a batch $\mathcal{B}$, the overall training objective is as Equation \eqref{eq: overall loss}.
\begin{equation} \label{eq: overall loss}
    \mathcal{L} = \sum_{\hat{\mathcal{T}}^i\in \mathcal{B}} (\mathcal{L}_{FC}^i + \mathcal{L}_{AC}^i + \mathcal{L}_{FC'}^{i} + \mathcal{L}_{AC'}^{i})
\end{equation}

\subsubsection{Model Training and Adaptation}
In our work, we maintain a queue of target entities encoded from preceding mini-batches \cite{he2020MoCo}. 
For each batch, we enqueue the current batch of entities and dequeue the oldest.
For a particular sample, the samples in the queue except itself are seen as negative samples, including in-batch and preceding-batch samples.
In this way, the negative samples are decoupled from the mini-batch size and the encoder learns to distinguish the ground truth entity from more negative entities.
Moreover, we filter out the triples with the same ground truth entity with masks to avoid contradiction. 
In implementation, we calculate the negative logits as $s^n=exp(\mathbf{q}\cdot \mathbf{e}^{queue}/\tau)$ where $\mathbf{e}^{queue}$ are the queue of preceding batch entity embeddings. 
Then we mask the positive and negative logits of the non-negative entities.
The training procedure is illustrated in detail in Algorithm \ref{alg: pipeline}. 

For parameter-efficient pre-trained model adaptation, we fix the parameter of the visual encoder and only fine-tune the parameters of VMN and the PLMs of the query encoder and entity encoder.
The efficient adaptation of the pre-trained vision model makes the trainable parameter size of our framework similar to PLM-based methods. 
We explore the effect of different PLMs in Section \ref{sec: plms}.

\subsection{Knowledge Representation Memorization}

We propose to memorize the knowledge representations in a Knowledge Store (KS) as Equation \eqref{eq: KS} with query embeddings as keys and their entity labels as values.
\begin{equation} \label{eq: KS}
    (\mathcal{Q}, \mathcal{E}) = \{(f_q(h,r), e)|(h,r,e)\in \mathcal{T} \}
\end{equation}

We also memorize entity embeddings and entity labels in the Entity Store (ES) as Equation \eqref{eq: ES}. 
\begin{equation} \label{eq: ES}
    (\mathcal{E}, \mathcal{E}) = \{(f_e(e), e)| e\in \mathcal{E}\}
\end{equation}

During the store construction process, the query encoder and entity encoder are both fixed without further training.

\textbf{\textit{Remark 2}}: \textit{
Here we analyze how the contrastive learning of CMR improves the performance of semantic neighbor retrieval.
In our work, the similarity between the query and its target entity embedding is enlarged after contrastive learning.
Consequently, the queries with the same target entity should also be similar.
That is, if a query $q = (h,r,?)$ and $q'=(h',r',?)$ share the same target entity $t$, $s(q,q')$ should be large since both $s(q,t)$ and $s(q',t)$ are large.
Thus, if we retrieve the semantic neighbor $q''$ of $q$ that $s(q,q'')$ is large, the target entity of $q''$ is highly probable to be $t$ and helps with the inference of $q$. 
Thus our proposed contrastive learning is crucial for the retrieved semantic neighbors to help the inference.
}

\begin{algorithm}[!t]
    \SetKwInOut{Input}{Input}
    \SetKwInOut{Output}{Output}
    \SetKwInOut{Require}{Require}
    
    \Input{MKG $\mathcal{G}=\{\mathcal{E},\mathcal{R},\mathcal{T}\}$, 
    Train Triples $\hat{\mathcal{T}}$, Batch $\mathcal{B}$, queue of entities in preceding batches $\mathcal{Q}$ and $\mathcal{Q}_v$}
    \For{$\mathcal{B}=\{(h,r,t)\} \in loader(\hat{\mathcal{T}})$}{
        Calculate query embedding $\mathbf{q}$ of $(h,r)$ as Equation \eqref{eq: query emb}\\
        Calculate visual prefixes $\mathbf{V}$ of entity $t$ as Equation \eqref{eq: vmn}\\
        Calculate token embeddings $\mathbf{D}$ of entity $t$ \\
        Calculate entity embedding $\mathbf{e}_f$ as Equation \eqref{eq: ent encoder}\\
        Enqueue$(\mathcal{Q},\mathbf{e}_f)$, Dequeue$(\mathcal{Q})$ \\
        Enqueue$(\mathcal{Q}_v,\bar{\mathbf{V}})$, Dequeue$(\mathcal{Q}_v)$ \\
        Calculate positive logits $s_f$ and $s_p$  \\
        Calculate negative logits $s^n_f$ and $s^n_p$\\
        \For{$(h^i,r^i,t^i)\in \mathcal{B}$}{
        Construct the negative sample mask $M$ from $\mathcal{N}^i$ as
        $\mathcal{N}^i = \{(h^i,r^i,n)|n \in \mathcal{Q},(h^i,r^i, n) \notin \hat{\mathcal{T}} \}$\\
        }
        Mask the negative logits with $M$ \\
        Optimize the model with overall loss as Equation \eqref{eq: overall loss}\\
        }

    \caption{Contrastive learning procedure in CMR}
    \label{alg: pipeline}
\end{algorithm}

\subsection{Retrieval-augmented Inductive Inference}
For inductive inference, given the query $q=(h,r,?)$, we first get the query embedding $f_q(h,r)$. Then we get two output distributions over all the entities $p_{KS}$ and $p_{ES}$. 

For $p_{KS}$, we query the KS with $f_q(h,r)$ to retrieve its $k$-nearest semantic neighbors $\mathcal{SN}$ according to the representation distance. We implement with L2 distance retrieval with FAISS \cite{johnson2019faiss}, which is equal to cosine similarity search.
Then, we apply softmax on the negative distances and aggregate probability over all the entities in the entity set $\mathcal{E}$ as Equation \eqref{eq: p_ks}. 
The entities that are not in the $\mathcal{SN}$ have zero probability. 
In the implementation, we only contain one nearest semantic neighbor in $p_{KS}$ for each target $e_i$ \cite{khandelwal2019KNNLM,zhang2022kNNKGE}. 
\begin{equation} \label{eq: p_ks}
    p_{KS}(y|(h,r)) \propto  \sum_{(\mathbf{q}_i, e_i)\in \mathcal{SN}} \mathds{1}_{y=e_i} \exp(-d(f_q(h,r), \mathbf{q}_i))
\end{equation}

For $p_{ES}$, we retrieve all the entities in the ES and rank them according to their similarity to the query. 
We apply softmax on the similarity distribution to obtain the $p_{ES}$ as Equation \eqref{eq: p_ES}.  
\begin{equation} \label{eq: p_ES}
    p_{ES}(y|(h,r)) \propto \exp(s(f_q(h,r), f_e(e_i))), e_i \in ES
\end{equation}

The final output distribution is the linear interpolation of $p_{ES}$ with $p_{KS}$ with a hyper-parameter $\lambda$ as Equation \eqref{eq: output}.
\begin{equation} \label{eq: output}
    p(y|(h,r)) = \lambda p_{KS}(y|(h,r)) + (1-\lambda)p_{ES}(y|(h,r))
\end{equation}

\begin{table}[!b]
    \caption{Dataset statistics of inductive MKGC.}
        \vspace{-12pt}
    \begin{center}
    \resizebox{0.48\textwidth}{!}{
    \small
    \begin{tabular}{lrrrrrrr}
    \toprule
    & \multirow{2}{*}{$|\mathcal{R}|$} &  \multicolumn{2}{c}{Train} & \multicolumn{2}{c}{Valid} & \multicolumn{2}{c}{Test}  \\
    \cmidrule{3-8}
    & & $|\mathcal{E}|$ & $|\mathcal{T}|$ & $|\mathcal{E}|$ & $|\mathcal{T}|$ & $|\mathcal{E}|$ & $|\mathcal{T}|$ \\
    \midrule
    FB15K-237 ind & 237 & 11,633 & 215,082 & 1,454 & 42,164 & 1,454 & 52,870 \\
    WN18RR ind & 11 & 32,755 & 69,585 & 4,094 & 11,381 & 4094 & 12,087 \\
    WN9 ind & 9 & 5,245 & 11,357 & 655 & 1,512 & 655 & 1,528 \\
    \bottomrule
    \end{tabular}}
    \end{center}
        \vspace{-10pt}
    \label{tab: datasets} 
    \end{table}
\begin{table}[!b]
    \caption{Dataset statistics of transductive MKGC.}
\vspace{-12pt}
    \begin{center}
    
    \resizebox{0.4\textwidth}{!}{
    \small
\begin{tabular}{lrrrrr}
\toprule
Dataset   & $|\mathcal{R}|$ & $|\mathcal{E}|$ & Train & Valid & Test \\
\midrule
FB15K-237 & 237    & 14,541 & 272,115 & 17,535  & 20466  \\
WN18      & 18     & 40,943 & 141,442 & 5,000   & 5,000  \\
WN9       & 9      & 6,555  & 11,741  & 1,337   & 1,319 \\
\bottomrule
    
    \end{tabular}}
    \end{center}
    \label{tab: trans_datasets} 
    \end{table}

\begin{table*}[!t]
\caption{Results of inductive multimodal knowledge graph completion. The \textbf{best} and \underline{second best} results of each column are highlighted. 
The results with $^*$ are based on our re-implementation with their source code. Other results are from the original paper.
``-text'' means only text modality is used.
The $\mathcal{L}_{AC}$ denotes the pre-align contrastive loss. $\mathbf{D}$ denotes the description tokens in the entity encoder. The $p_{KS}$ and $p_{ES}$ denote the probability from knowledge store and entity store, respectively. 
}
\resizebox{\linewidth}{!}{
\centering
\small

\begin{tabular}{l cccc c cccc c cccc}
\toprule
\multirow{2}{*}{Model} & \multicolumn{4}{c}{FB15K-237 ind} & \makebox[0.005\textwidth][c]{} & \multicolumn{4}{c}{WN18RR ind} &
\makebox[0.005\textwidth][c]{} &
\multicolumn{4}{c}{WN9 ind$^*$} \\
\cmidrule{2-5}
\cmidrule{7-10}
\cmidrule{12-15}
& MRR & Hits@1 & Hits@3 & Hits@10        && MRR & Hits@1 & Hits@3 & Hits@10        && MRR & Hits@1 & Hits@3 & Hits@10        \\
       \midrule
GloVe-BOW & 0.172  & 0.099  & 0.188  & 0.316  &  & 0.170  & 0.055  & 0.215  & 0.405  &  & 0.301  & 0.168  & 0.366  & 0.558  \\
BE-BOW & 0.173  & 0.103  & 0.184  & 0.316  &  & 0.180  & 0.045  & 0.244  & 0.450  &  & 0.263  & 0.122  & 0.336  & 0.526  \\
GloVe-DKRL        & 0.112  & 0.062  & 0.111  & 0.211  &  & 0.115  & 0.031  & 0.141  & 0.282  &  & 0.191  & 0.085  & 0.228  & 0.402  \\
BE-DKRL   & 0.144  & 0.084  & 0.151  & 0.263  &  & 0.139  & 0.048  & 0.169  & 0.320  &  & 0.241  & 0.153  & 0.260  & 0.425  \\
BLP-TransE        & 0.195  & 0.113  & 0.213  & 0.363  &  & 0.285  & 0.135  & 0.361  & 0.580  &  & 0.434  & 0.314  & 0.496  & 0.672  \\
BLP-DistMult      & 0.146  & 0.076  & 0.156  & 0.286  &  & 0.248  & 0.135  & 0.288  & 0.481  &  & 0.208  & 0.079  & 0.240  & 0.498  \\
BLP-ComplEx       & 0.148  & 0.081  & 0.154  & 0.283  &  & 0.261  & 0.156  & 0.297  & 0.472  &  & 0.314  & 0.215  & 0.343  & 0.515  \\
BLP-SimplE        & 0.144  & 0.077  & 0.152  & 0.274  &  & 0.239  & 0.144  & 0.265  & 0.435  &  & 0.328  & 0.218  & 0.377  & 0.541  \\
MoSE-AI$^*$   & 0.105  & 0.077  & 0.114  & 0.156  &  & 0.056  & 0.033  & 0.058  & 0.098  &  & 0.143  & 0.073  & 0.160  & 0.271  \\
MoSE-BI$^*$   & 0.206  & 0.137  & 0.221  & 0.348  &  & 0.318  & 0.213  & 0.362  & 0.527  &  & 0.323  & 0.204  & 0.370  & 0.568  \\
MoSE-MI$^*$   & 0.228  & 0.137  & 0.254  & 0.410  &  & 0.335  & 0.228  & 0.379  & 0.554  &  & 0.364  & 0.245  & 0.408  & 0.615  \\
\midrule
StAR   & 0.163  & 0.092  & 0.176  & 0.309  &  & 0.321  & 0.192  & 0.381  & 0.576  &  & 0.209  & 0.094  & 0.239  & 0.462  \\
kNN-KGE   & 0.198  & 0.146  & 0.214  & 0.293  &  & 0.294  & 0.223  & 0.320  & 0.431  &  & -   & -   & -   & -   \\
StATIK & 0.224  & 0.143  & 0.248  & 0.381  &  & 0.516  & 0.425  & 0.558  & 0.690  &  & -   & -   & -   & -   \\
CMR-text& \underline{0.349} & \underline{0.259} & \underline{0.396} & \underline{0.523} & & \underline{0.608} & \underline{0.534} & \underline{0.632} & \underline{0.779} && \underline{0.858} & \underline{0.811} & \underline{0.897} & \textbf{0.939} \\
\textbf{CMR} & \textbf{0.371} & \textbf{0.269} & \textbf{0.418} & \textbf{0.567} &  & \textbf{0.631} & \textbf{0.577} & \textbf{0.644} & \textbf{0.788} &  & \textbf{0.879} & \textbf{0.848} & \textbf{0.898} & \underline{0.932}\\
\midrule
~~w/o $\mathcal{L}_{AC}$    & 0.356  & 0.263  &  0.395  & 0.541 && 0.601 & 0.542 & 0.626 & 0.779 && 0.858 & 0.813 & 0.886 & 0.932  \\
~~w/o $\mathbf{D}$ & 0.298 & 0.215 & 0.338 & 0.466 && 0.531 & 0.503 & 0.547 & 0.577 && 0.842& 0.835&0.843 & 0.853   \\
~~w/o $p_{ES}$ & 0.341 & 0.235 & 0.390 & 0.553 && 0.582 & 0.549 & 0.596 & 0.642 && 0.818 & 0.818 & 0.818 & 0.818 \\
~~w/o $p_{KS}$    & 0.258  & 0.177  &  0.281  & 0.422 && 0.424 & 0.300 & 0.489 & 0.664 && 0.509 & 0.386 & 0.579 & 0.740 \\ 
\bottomrule
\vspace{-10pt}
\end{tabular}}
\label{tab: main_result}
\end{table*}

\section{Experiments}

\subsection{Experimental Setting}

\subsubsection{Datasets.}
We evaluate our proposed method on three inductive multimodal KGC datasets. 
For the inductive datasets, we augment the existing IKGC datasets \cite{daza2021blp} \textbf{FB15K-237 ind} and \textbf{WN18RR ind} with entity images following previous studies \cite{zhao2022mose,chen2022mkgformer} and entity descriptions \cite{daza2021blp,yao2019kgbert}. 
Moreover, we follow \citeauthor{daza2021blp} to transform an existing MKGC dataset {WN9} \cite{xie2017IKRL} to the inductive setting as \textbf{WN9 ind}. 
The datasets split the entity set $\mathcal{E}$ and the triple set $\mathcal{T}$ into disjoined train/valid/test sets. 
The train graph consists of triples with entities only in the $\mathcal{E}_{train}$. 
The valid or test triples consist of triples with the entities in $\mathcal{E}_{valid}\cup\mathcal{E}_{train}$ or $\mathcal{E}_{test}\cup\mathcal{E}_{valid}\cup\mathcal{E}_{train}$ respectively \cite{daza2021blp, markowitz2022statik}.
The valid and test triples have unseen entities at the position of the head, tail, or both \cite{daza2021blp}. The dataset statistics are shown in Table \ref{tab: datasets}.

We also conducted experiments on three widely used transductive MKGC datasets: \textbf{FB15K-237}, \textbf{WN18}, and \textbf{WN9}.
Under the transductive setting, all the entities in the test set appear in the training set. 
The dataset statistics are shown in Table \ref{tab: trans_datasets}.

\subsubsection{Evaluation.}
We evaluate each test triple with head entity prediction and tail entity prediction \cite{bordes2013TransE} and calculate the rank of the target entity compared to all the candidate entities from both seen and unseen entities.
Four metrics are used for evaluation: Mean Reciprocal Rank (MRR), the mean of the reciprocal of the target entity, and Hits@K (K=1,3,10), the accuracy in the top K predicted entities.
Higher MRR and higher Hits@K indicate better performance.
We adopt the filtered setting \cite{bordes2013TransE} to exclude other target entities of the same query when calculating evaluation metrics.

\subsubsection{Implementation details.}
We implement our framework with PyTorch and Transformers \cite{wolf2020huggingface}.
We exploit BERT \cite{devlin2019bert} as query encoder and entity encoder that does not share parameters, and exploit ViT \cite{dosovitskiy2020ViT} as visual encoder.
All entities have description information and the entities without images are padded with random values.
The visual prefix length is 4 for all datasets. The total sequence length including prefixes is 50. The temperature $\tau$ is 0.05. The mini-batch size is 768 and we maintain a queue size of 3 mini-batches, consisting of 2 preceding batches and 1 current batch. We use AdamW optimizer with linear learning rate decay. The training is early stopped after 3 epochs without improvements in Hits@1 on the validation set. 
At test time, we fixed the encoders and memorized all the triples with both seen and unseen entities in KS. 
During semantic neighbor retrieval, we filter out the semantic neighbors with the same head and relation to prevent information leaks.
We searched the hyperparameter $\lambda$ and $k$ in the validation set and selected the ones with the best results.
We ran our experiments on 1 A6000 GPU (48GB RAM).

\subsubsection{Baselines.}
For the inductive baselines, we employ two groups of baselines: (1) KGE-based methods:  \textbf{DKRL} \cite{xie2016DKRL}, \textbf{BLP} \cite{daza2021blp}, \textbf{MoSE} \cite{zhao2022mose}, (2) PLM-based methods: \textbf{StAR} \cite{wang2021StAR}, \textbf{$k$NN-KGE} \cite{zhang2022kNNKGE}, \textbf{StATIK} \cite{markowitz2022statik}. 
In the baselines, the IKGC baselines usually model two modalities: structure and text, including DKRL, BLP, StAR, kNN-KGE, and StATIK, while the MKGC baseline models three modalities: structure, text, and visual, including MoSE.
Since there is no inductive MKGC methods, we extend the MKGC method MoSE to the inductive setting by utilizing text and visual embeddings and randomly initializing their structure embeddings. 
We did not include the VLM-based MKGC methods \cite{chen2022mkgformer,liang2023SGMPT} because they could not directly transfer to the inductive setting without re-training. 
We compare the baselines with our proposed CMR and the variant without visual modality, CMR-text, for a fair comparison with the textual IKGC baselines.

For the transductive baselines, we compare CMR with two groups of representative MKGC baselines: (1) KGE-based methods: \textbf{TransAE} \cite{wang2019transAE}, \textbf{RSME} \cite{wang2021rsme}, \textbf{MoSE} \cite{zhao2022mose}, and (2) VLM-based methods: \textbf{MKGformer} \cite{chen2022mkgformer}, \textbf{SGMPT} \cite{chen2022mkgformer}.

\subsection{Main Results}
As shown in Table \ref{tab: main_result}, CMR outperforms all the baselines on three inductive datasets, FB15K-237 ind, WN18RR ind, and WN9 ind. 

Compared to KGE-based IKGC methods \cite{xie2016DKRL,daza2021blp}, CMR achieves 17.6\%, 3.6\%, and 42.4\% improvements in MRR compared to BLP-TransE on three datasets respectively. 
Compared to MKGC methods \cite{zhao2022mose}, CMR achieves 14.3\%, 29.6\%, and 49.4\% improvements in MRR compared with MoSE-MI on three datasets respectively. 
The reason could be that the structural information from KGE scores without re-training is limited for generalizing to unseen entities.

Compared to PLM-based methods \cite{wang2021StAR,zhang2022kNNKGE,markowitz2022statik}, CMR achieves 14.7\% and 11.5\% improvements in MRR on FB15K-237 ind and on WN18RR ind compared with state-of-the-art baseline StATIK.
Moreover, the textual version CMR-text outperforms StATIK by 12.5\% and 9.2\% on FB15K-237 ind and WN18RR ind.
Moreover, CMR and CMR-text both outperform kNN-KGE, which utilizes retrieval-augmented MLM inference. The reason could be that the CMR has better generalization ability since our proposed contrastive learning directly optimizes the query-entity similarity and leads the semantic neighbors with the same ground truth entity to be closer. Thus the performance of retrieved semantic neighbors from CMR is boosted.

Compared with CMR-text, CMR achieves 2.2\%, 2.3\%, and 18.3\% improvements in MRR on three datasets, demonstrating that visual modality is helpful for IKGC. 
Moreover, the addition of visual modality does not interfere with the performance of CMR-text, which shows that CMR could alleviate the modality contradiction.
The improvements on WN9 are relatively large because (1) the image are high-qualified \cite{xie2017IKRL}, and (2) CMR took advantage of the unremoved reversed triples like other two datasets \cite{toutanova2015fb237,dettmers2018conve-wn18rr}.

\begin{table*}[!t]
\caption{Results of transductive multimodal knowledge graph completion. The \textbf{best} and \underline{second best} results of each column are highlighted. The results of the baselines are from their original paper.
}
\vspace{-10pt}
\resizebox{\linewidth}{!}{
\centering
\small
\begin{tabular}{l cccc c cccc c cccc}
\toprule
\multirow{2}{*}{Model} & \multicolumn{4}{c}{FB15K-237} & \makebox[0.005\textwidth][c]{} & \multicolumn{4}{c}{WN18} &
\makebox[0.005\textwidth][c]{} &
\multicolumn{4}{c}{WN9} \\
\cmidrule{2-5}
\cmidrule{7-10}
\cmidrule{12-15}
&MR        & Hits@1 & Hits@3 & Hits@10 & &MR        & Hits@1 & Hits@3 & Hits@10 && MR   &Hits@1 & Hits@3 & Hits@10  \\
       \midrule
TransAE &431 & 0.199 & 0.317 & 0.463 &  &352 & 0.323 & 0.835 & 0.934 && 17  & - & - & 0.942  \\
RSME  & 417 & 0.242 & 0.344 & 0.467 & &223 & {0.943} & 0.951 & 0.957 &  & 55& 0.878 & 0.912 & 0.923  \\
MoSE-MI & \underline{127} & \textbf{0.268} & \underline{0.394} & \underline{0.540} &  & \textbf{7} & \underline{0.948} & {0.962} & {0.974} && \underline{4} & \underline{0.909} & \underline{0.937} & \underline{0.967}  \\
\midrule
MKGformer & 221 & 0.256 & 0.367 & 0.504 & & {28} &  0.944 & 0.961 & 0.972 & & - & - & - & -  \\
SGMPT& 238  & {0.252}  & {0.370} &  {0.510}  & & 29 & {0.943} & \underline{0.966} & \underline{0.978}&& -& - & - & -  \\
CMR &\textbf{123} & \underline{0.263} & \textbf{0.395} & \textbf{0.543} & & \underline{17} &\textbf{0.993} &\textbf{0.995}&\textbf{0.997}&& \textbf{1}& \textbf{0.916} & \textbf{0.983} & \textbf{0.997}   \\
\bottomrule
\end{tabular}}
\label{tab: trans_result}
\end{table*}

\subsection{Ablation Study}
The ablation study is shown in Table \ref{tab: main_result}. 

To evaluate the effectiveness of our proposed contrastive learning, we ablate the $\mathcal{L}_{AC}$ and $\mathbf{D}$ in the entity embeddings. 
The performance of CMR decreases without $\mathcal{L}_{AC}$ on three datasets. 
The reason could be that the pre-alignment of queries and VMN-mapped images increases the generalization of VMN. 
Moreover, the textual-visual alignment of $\mathcal{L}_{AC}$ further improves the textual-multimodal alignment of $\mathcal{L}_{FC}$.
CMR w/o $\mathbf{D}$ shows the results from only textual-visual contrastive learning and outperforms most IMKGC baselines.
It indicates that the textual-visual correlation could help the IMKGC. 
However, CMR-text outperforms CMR w/o $\mathbf{D}$, where the former captures textual-textual correlation.
It demonstrates that the entity descriptions provide more helpful information than entity images. 
Moreover, CMR outperforms CMR-text and CMR w/o $\mathbf{D}$, showing that CMR could effectively capture multimodal correlations.

To evaluate the effectiveness of the semantic neighbor retrieval-enhanced inference, we separately report the results of $p_{KS}$ and $p_{ES}$.
We report the best performance of CMR w/o $p_{ES}$, i.e. $p_{KS}$, throughout the number of semantic neighbors $k$. 
The CMR w/o $p_{KS}$ and CMR w/o $p_{ES}$ both underperform CMR, demonstrating the necessity of both semantic neighbors and direct query-entity similarity measures. 
The reason for best $p_{KS}$ outperforms $p_{ES}$ could be that semantic neighbors aggregate more helpful information than the triple itself. 
Moreover, CMR w/p $p_{KS}$ outperforms the baselines on FB15K-237 ind and WN9 ind, demonstrating that the unified cross-modal contrastive learning effectively utilizes multimodal information for inductive MKGC.

\begin{figure}[!t]
\centering
\includegraphics[width=0.49\textwidth]{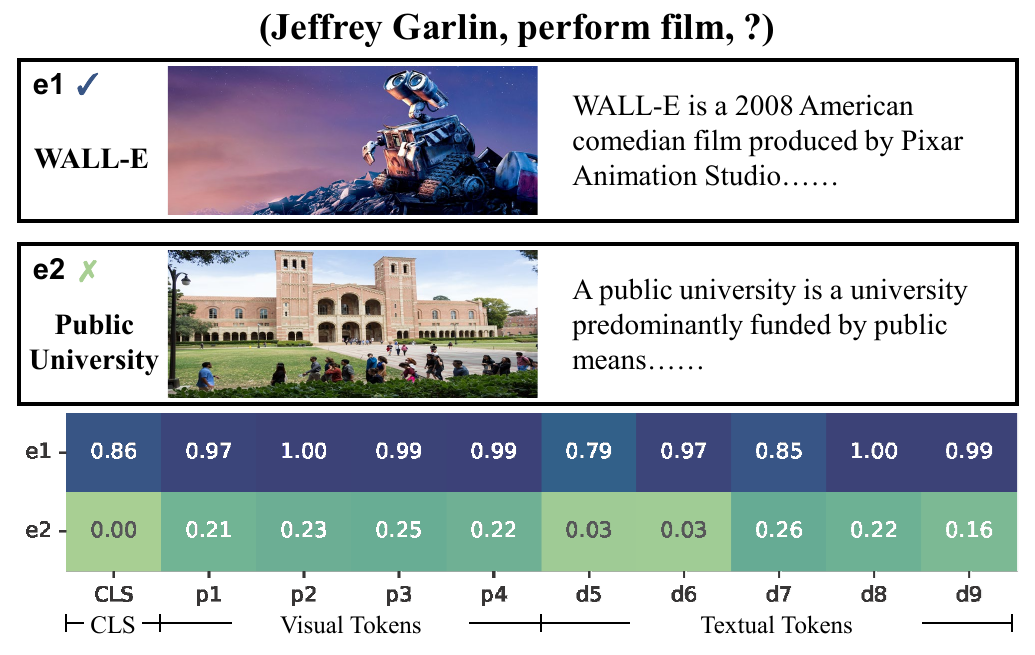}
\caption{Token-wise similarity of query and ground-truth entity e1 versus query and irrelevant entity e2.}\label{fig:heatmap}
\vspace{-10pt}
\end{figure}

\subsection{Transductive Results}

We conducted experiments under the transductive setting to evaluate the effectiveness of CMR in Table \ref{tab: trans_result}. 
The transductive MKGC could also benefit from the generalization ability of CMR since there are always unseen or less-seen entities in KGs \cite{tan2023kracl}. 

CMR reaches the best on most metrics compared with baselines on three transductive datasets. 
Compared with the KGE-based methods, CMR outperforms the SOTA method MoSE-MI in most metrics. Compared with VLM-based methods, CMR outperforms all the baselines. 
The reason could be that our contrastive learning could avoid modality contradiction and the semantic neighbors could help the entities with limited structural information. 
The transductive MKGC could also benefit from the generalization ability of CMR since there are always unseen or less-seen entities in KGs \cite{tan2023kracl}. 
CMR significantly outperforms the baselines on WN18 and WN9 datasets. The reason could be that the WN18 and WN9 also have unremoved triples problems, thus the semantic neighbor retrieval could help predict this kind of triples.

\begin{table}[t!]
\caption{The top-3 prediction of $p_{ES}$, $\mathcal{SN}$ of the query, $p_{KS}$ and CMR with unseen entity \textit{WALL-E} as query and target respectively. True target entities are \textbf{boldfaced}.
}
\vspace{-10pt}
\small
\centering
\resizebox{0.46\textwidth}{!}{
\begin{tabular}[t]{c|p{3cm}|p{3cm}}
\toprule
 & \textbf{As query}     & \textbf{As target}\\
\midrule
Query & (\textit{WALL-E}, genre) & (Pixar, nominated) \\
\midrule
\multirow{3}{*}{$p_{ES}$} & 1. Humour & 1. A Bug's Life \\
& 2. Screwball comedy & 2. Toy Story 2 \\
& 3. Comedy& 3. Ratatouille \\
\midrule
\multirow{3}{*}{$\mathcal{SN}$}  
& 1. (Lilo \& Stitch, genre)  & 1. (Pixar, product$^{-1}$) \\
& 2. (Shrek 2, genre) & 2. (Pixar, product$^{-1}$) \\
& 3. (Shrek, genre) & 3. (Stanton, nominated) \\
\midrule
\multirow{3}{*}{$p_{KS}$}  
& 1. \textbf{Science Fiction}  & 1. \textbf{\textit{WALL-E}} \\
& 2. Fairy tale & 2. Toy Story 3 \\
& 3. Costume drama & 3. Ratatouille \\
\midrule
\multirow{3}{*}{CMR}  & 1. \textbf{Science Fiction} & 1. \textbf{\textit{WALL-E}} \\
& 2. Fairy tale & 2. A Bug's Life \\
& 3. Costume drama & 3. Toy Story 2 \\
\bottomrule
\end{tabular}
}
\vspace{-15pt}
\label{tab: case}
\end{table}

\subsection{Case Study}
\subsubsection{Inductive cross-modal correlations.}
We visualize the token-wise similarity between query embedding and the token embeddings of the ground-truth entity and irrelevant entity in Figure \ref{fig:heatmap}, where both entities are unseen during training. 
The similarity of visual tokens $p_1-p_4$ and textual tokens $d_5-d_9$ of ground-truth entity $e_1$ are all high while those of the irrelevant entity $e_2$ are all low.
It shows that CMR could dynamically capture the textual-textual and textual-visual correlation between the query and multimodal tokens of entities.
It also indicates that CMR could distinguish the relevant entity from the irrelevant ones by effective unified cross-modal contrastive learning.

\subsubsection{Semantic neighbors Retrieval}
As shown in Table \ref{tab: case}, we conducted a case study to explore how explicit memorization and semantic neighbor retrieval helps inductive inference with unseen entities as query and as target, respectively.

\textbf{With an unseen entity in the query}, the retrieval of semantic neighbors finds the nearest similar queries in KS that could imply the true target entity label. 
For example, like \textit{WALL-E}, \textit{Lilo \& Stitch} is also a film about space and they both belong to \textit{Science Fiction} genre. 
Moreover, the semantic neighbors are not limited by the topology linkage, since \textit{Lilo \& Stitch} is not directly linked to \textit{WALL-E}.

\textbf{With an unseen entity as the target}, the semantic neighbors could also imply the true target label. 
For example, the films \textit{Pixar} nominated for are usually the films produced by \textit{Pixar}. 
Thus the \textit{(Pixar, product$^{-1}$, WALL-E)} in the test graph provide the true target entity of (\textit{Pixar, nominated}).
It also shows that the structural neighbors are implicitly modeled by CMR when they are also semantic neighbors of the query.

\begin{table}[!t]
\caption{Results of CMR on WN9 ind with different PLMs. }\label{tab: PLMs}
\vspace{-10pt}
\begin{center}
\resizebox{\linewidth}{!}{
\centering
\small
\begin{tabular}{llcccc}
\toprule
\multirow{2}{*}{Model} &\multirow{2}{*}{Paras} & \multicolumn{4}{c}{WN9 ind}  \\
\cmidrule{3-6}
&& MRR & Hits@1 & Hits@3 & Hits@10 \\
       \midrule
CMR-BERT$_{base}$& 220.9M & {0.858} & {0.813} & {0.886} & {0.932}  \\
CMR-CLIP$_{base}$& 151.3M  & 0.862 & 0.825 & 0.890 & 0.926 \\
CMR-RoBERTa$_{base}$& 251.3M & 0.869 & 0.827 & 0.899 & 0.934 \\

\bottomrule
\end{tabular}}
\end{center}
\vspace{-10pt}
\end{table}

\subsection{Effect of Pretrained Encoders} \label{sec: plms}
We select BERT-base as our query encoder and entity encoder for a fair comparison with the BERT-based baselines.
To explore the ability of CMR with different pretrained encoders, we conducted experiments by replacing the BERT with widely-used PLMs,  CLIP-base \cite{radford2021CLIP} and RoBERTa-base \cite{liu2019roberta} in Table \ref{tab: PLMs}.
With BERT and RoBERTa, the ViT remains fixed and only the BERT/RoBERTa and the VMN are trained, while with CLIP we finetuned both the visual and textual encoder of CLIP considering its low parameter.
As shown in Table \ref{tab: PLMs}, the performance of different PLMs varies in a certain range on WN9 ind. 
The improvements from CLIP$_{base}$ could be attributed to the visual-textual contrastive pretraining, and that from RoBERTa$_{base}$ could be attributed to robust pretraining. 

\begin{figure}[t]
\centering
\vspace{-5pt}
\subfigure{
\label{fig:knn_k}
\includegraphics[width=0.37\textwidth]{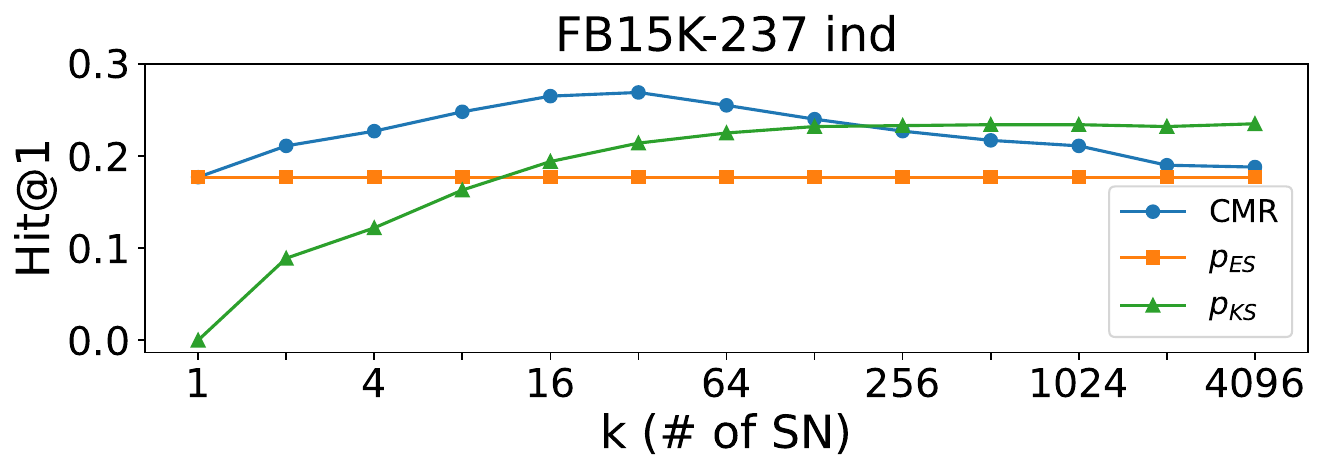}
}
\vspace{-10pt}
\subfigure{
\label{fig:knn_lambda}
\includegraphics[width=0.37\textwidth]{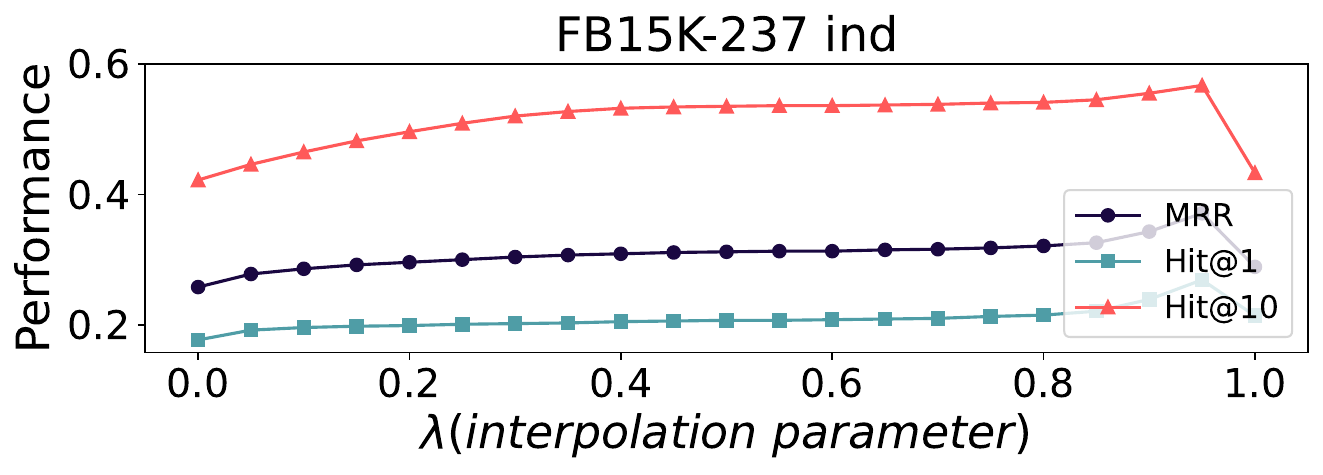}}
\caption{Performance with different semantic neighbor count $k$ and interpolation parameter $\lambda$ on FB15K-237 ind.} \label{fig:knn_paras}
\vspace{-10pt}
\end{figure}


\subsection{Effect of Retrieval Parameters} 
We conduct semantic neighbor retrieval parameter analysis of $k$ and $\lambda$ on FB15K-237 ind as shown in Figure \ref{fig:knn_paras}. 

\subsubsection{Parameter $k$}
The parameter $k$ controls the number of semantic neighbors retrieved and aggregated to the final prediction.
As shown in Figure \ref{fig:knn_k}, the performance of $p_{KS}$ tends to increase and converge to the best performance with more semantic neighbors. 
The reason could be that with more semantic neighbors, the included informative triples tend to converge to the best performance. 
The performance of CMR first increases then decreases, and finally converges to the performance of $p_{ES}$. 
The reason could be that with more semantic neighbors, the number of noisy triples also increases, thus the combination of $p_{ES}$ and $p_{KS}$ tends to decrease. 
However, the CMR consistently outperforms the $p_{ES}$, indicating the effective improvements brought by the aggregation of semantic neighbors.
Moreover, by adjusting the number of retrieved semantic neighbors, CMR could get the best performance and control the negative impact of the noisy semantic neighbors.

\subsubsection{Parameter $\lambda$}
The parameter $\lambda$ controls the interpolation ratio of $p_{KS}$ to the $p_{ES}$ according to Equation \ref{eq: output}.
We set $k=32$ to get the best performance of CMR according to Figure \ref{fig:knn_k} and conducted experiments with various $\lambda$.
As shown in Figure \ref{fig:knn_lambda}, with larger $\lambda$, the performance of CMR first increases. 
After reaching the best performance when $\lambda=0.95$, the performance decreases. 
It shows that by choosing the appropriate interpolation ratio of the semantic neighbor retrieval distribution, CMR could adjust the combination of both distributions and reach the best performance.

\begin{figure}[!t]
\centering
\includegraphics[width=0.27\textwidth]{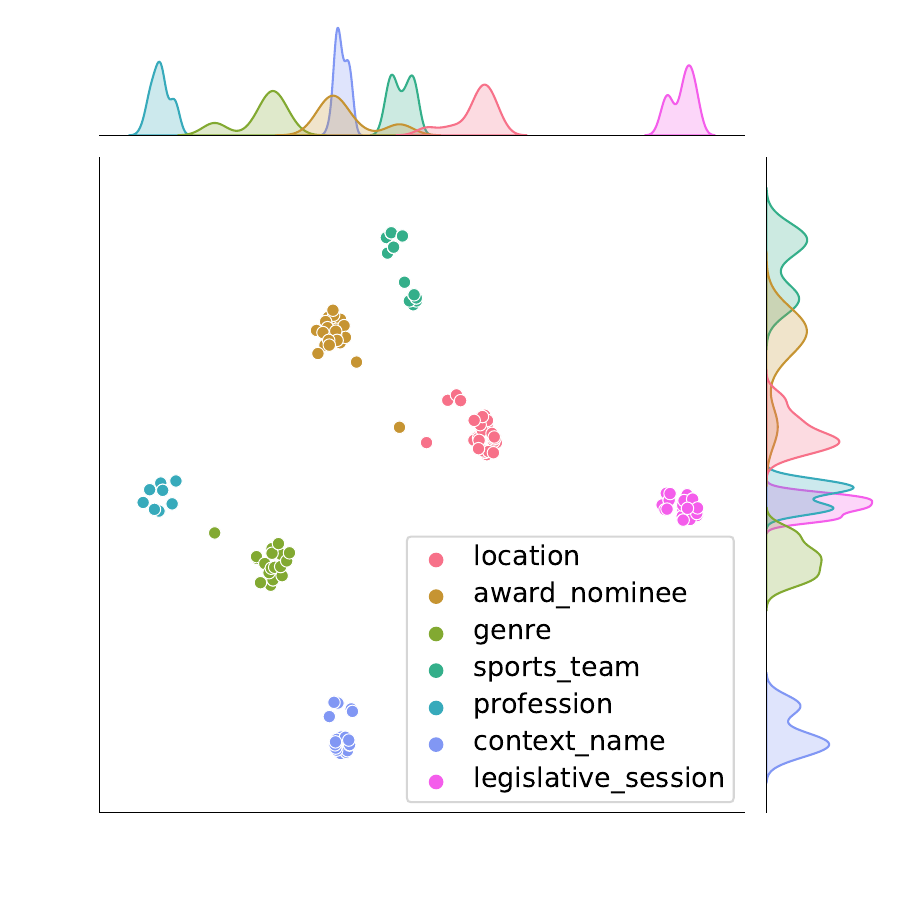}
\caption{The t-SNE \cite{van2008tsne} visualization of unseen entity embeddings with different entity types.}\label{fig:tsne}
\vspace{-10pt}
\end{figure}

\subsection{Visualization}
We show the t-SNE visualization of unseen entities in Figure \ref{fig:tsne}. We visualize the representations of unseen entities with different entity types (colors) to demonstrate different semantic information.
It shows that CMR could encode the entities with similar semantics closer in representation space even without being seen during training.
It indicates our generalizable representation ability.

\section{Conclusion}
In this paper, we propose to study the IMKGC task to address the emerging unseen entity problem in MKGs. 
We propose a semantic neighbor retrieval-enhanced IMKGC framework, CMR.
Specifically, we propose unified cross-modal contrastive learning to capture multimodal semantic correlation in a unified representation space and lead the helpful semantic neighbors closer, explicit knowledge representation memorization to support semantic neighbor aggregation, and semantic neighbor retrieval-enhanced inference to augment the query-entity similarity prediction.
Experiments validate the effectiveness and generalization ability of CMR on both inductive and transductive MKGC datasets. 
Future work directions include efficient semantic neighbor training, pretrained VLM exploration, and retrieval strategy enhancement.

\section*{Acknowledgements}
This research is supported by the National Natural Science Foundation of China (No. 62272250, U22B2048, 62077031), and the Natural Science Foundation of Tianjin, China (No. 22JCJQJC00150, 22JCQNJC01580).


\bibliographystyle{ACM-Reference-Format}
\bibliography{mybib}


\end{document}